\documentclass[11pt]{cernrep}
\usepackage{epsfig}
\usepackage{graphicx}
\usepackage{cite,./mcite}

\newcommand{\geqsim}{\,\raisebox{-0.6ex}{$\buildrel > \over \sim$}\,}

\begin{document}
\title{Have we seen anything beyond (N)NLO DGLAP at HERA? }
\author{Amanda Cooper-Sarkar}
\institute{Oxford University}
\maketitle
\begin{abstract}
The evidence from HERA for parton saturation, and other low-$x$ effects beyond 
the conventional DGLAP formalism, is 
recalled and critically reviewed in the light of new data and analyses 
presented at the conference. 
\end{abstract}


In the mid-90's the original surprise of the HERA Neutral Current $e^+p$ 
scattering data was the strong rise of the structure 
function $F_2$ at  low-$x$. This was taken to imply a strong rise of the 
gluon density at low-$x$ which was widely interpreted as implying the 
possibility of gluon saturation and the need for non-linear terms in the 
parton evolution equations. Even somewhat more conservative interpretations 
suggested the need to go beyond the DGLAP formalism at small-$x$, 
resumming $ln(1/x)$ as in the BFKL formalism.

\begin{figure}[tbp]
\vspace{-1.0cm} 
\centerline{
\epsfig{figure=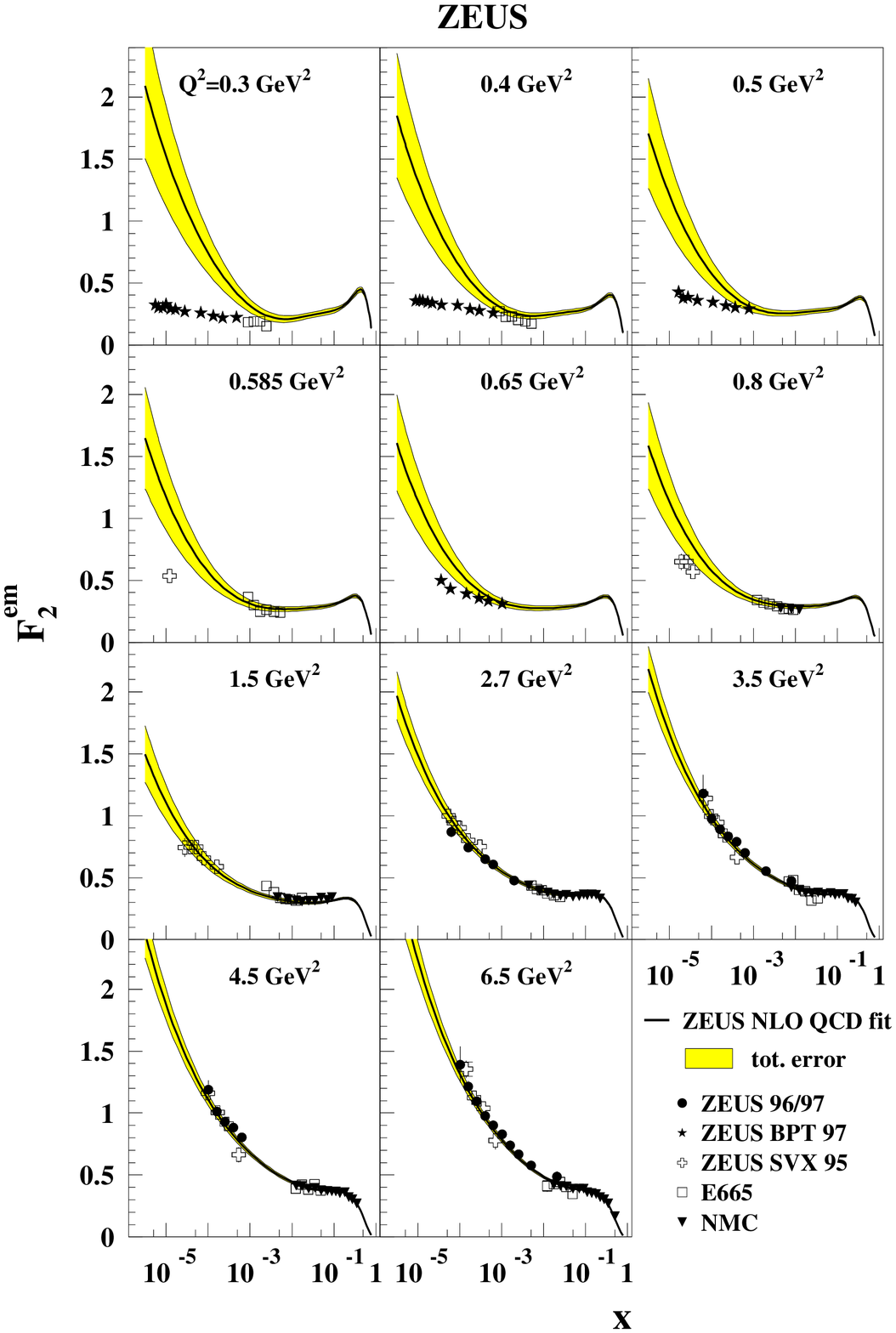,width=0.50\textwidth,height=9cm}
\epsfig{figure=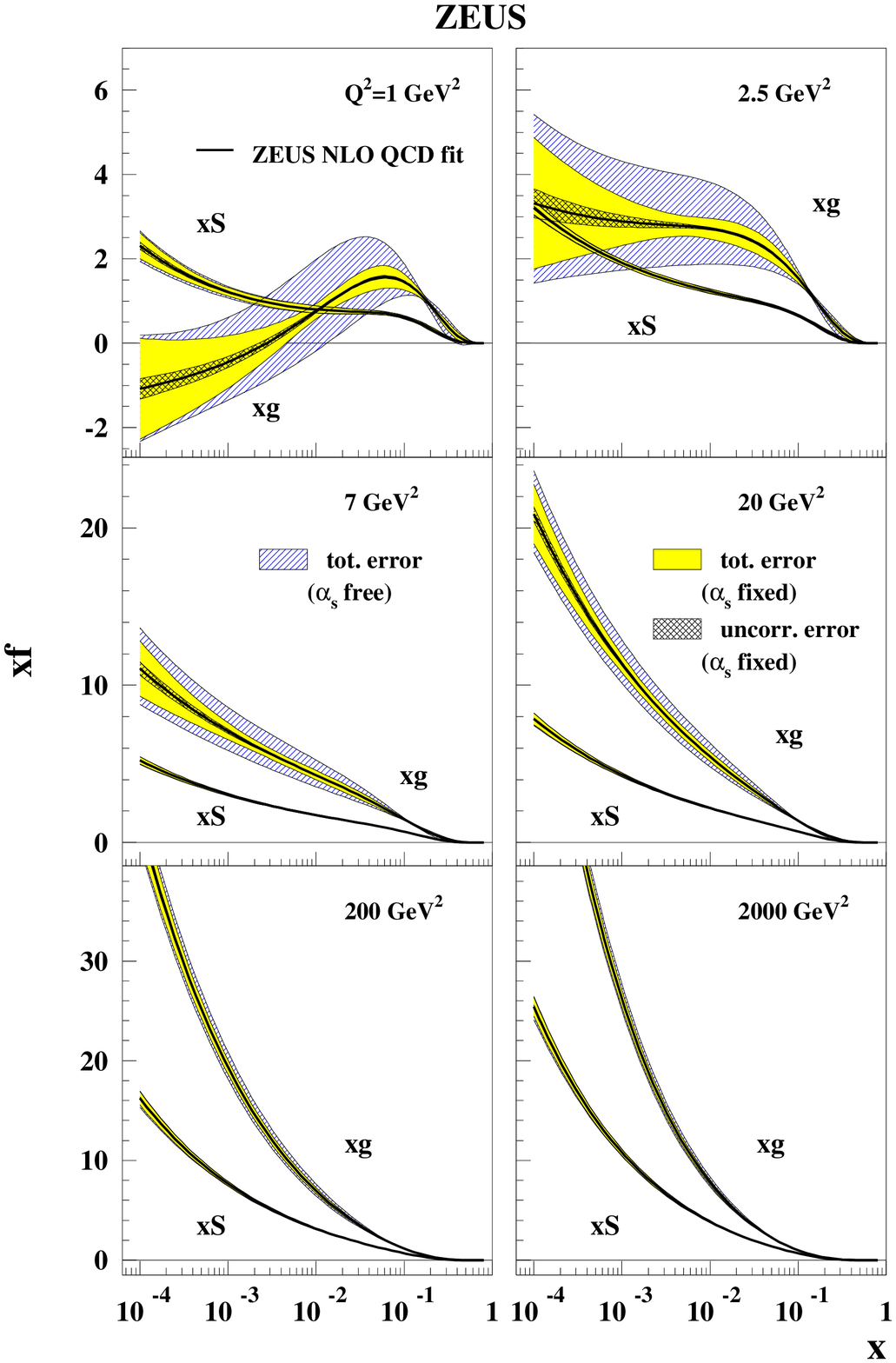,width=0.50\textwidth,height=9cm}
}
\caption {
Left plot: $F_2$ vs $x$ for various low $Q^2$ values.
Right plot: Sea and gluon PDF distributions extracted from a global PDF fit 
including these data. 
}
\label{fig:f2seaglu}
\end{figure}

However, at low-$x$ linear NLO DGLAP evolution itself predicts a rise in 
$F_2$, and 
in the gluon and sea PDFs, provided that $Q^2$ is large enough. 
One can begin parton evolution at a low $Q^2$ input scale, $Q^2_0$, 
using flat (or even valence-like) gluon and sea-quark input shapes in $x$ and 
the DGLAP $Q^2$ evolution will generate a steep low-$x$ rise 
of the gluon and sea at larger 
$Q^2 \gg Q^2_0$. The real surprise - seen in the data of the late 90's- 
was that steep shapes were already observed at 
rather low $Q^2$. Traditionally values of $Q^2_0\sim 4$GeV$^2$ were used, 
but the data already show a steep rise of $F_2$ at low-$x$ for  
$Q^2$ values, $Q^2 \sim 1$GeV$^2$, see Fig.~\ref{fig:f2seaglu} left-hand-side. 
To interpret these data in terms 
of conventional NLO DGLAP evolution we clearly 
need a low starting scale and thus we 
are forced into using perturbative QCD at a scale for which $\alpha_s(Q^2)$ 
is quite large- $\alpha_s(1.0) \sim 0.35$. 
Even if this is considered to be acceptable, we also need 
to use flexible input parton shapes, which can reproduce the steepness of the 
data. Surprisingly enough this does NOT imply that both the gluon and the sea 
input are already steep at $Q^2 \sim 1$GeV$^2$. The sea input is indeed steep, 
but the gluon input is valence-like, with a tendency to be negative at 
low-$x$!- see Fig.~\ref{fig:f2seaglu} right-hand-side.
(Essentially the gluon evolution must be fast in order that 
upward evolution can produce the extreme steepness of high-$Q^2$ data, 
however this also implies that downward evolution is fast and this results in 
the valence-like gluon at low-$Q^2$).

Thus when statements are made that HERA has established that the low-$x$ gluon 
is steep one must remember that this is only true for higher $Q^2$, 
$Q^2 \geqsim 10$GeV$^2$, within the DGLAP formalism.
However this formalism seems to work to much lower $Q^2$. Let us examine 
how the gluon and sea PDFs are extracted from the measurements. At low-$x$, 
the sea PDF 
is extracted fairly directly since, $F_2(x,Q^2) \sim x q(x,Q^2)$. 
 However the gluon PDF is 
extracted from the scaling violations, 
$\partial F_2/ \partial ln(Q^2) \sim P_{qg} xg(x,Q^2)$, 
such that the measurement 
is related to a convolution of the splitting function $P_{qg}$ and the 
gluon distribution. Thus  if the correct
splitting function is NOT that of the conventional DGLAP formalism, 
or if a more complex non-linear realtionship is needed, then a turn over of 
the data 
$\partial F_2/ \partial ln(Q^2)$ at 
low-$Q^2$ and low-$x$ may not imply a turn over of the gluon distribution.
It was suggested that measurements of other gluon related quantities could 
help to shed light on this question and the longitudinal structure function, 
$F_L$, and the heavy quark structure functions,
 $F_2^{c\bar{c}}, F_2^{b\bar{b}}$,
 are obvious candidates. All of these quantities have now been measured 
(see talks of K.~Papageorgiou and P.~Thompson in these proceedings) and,
 within present experimental uncertainties, they can be explained by the 
conventional NLO DGLAP formalism (with the heavy quark results shedding more 
light on the complexities of general-mass-variable-flavour number schemes 
than on the gluon PDF). 

Other measurements of more exclusive quantities can also give information on 
the correctness of the conventional formalism at low-$x$. 
For example HERA forward jet mesaurements 
(see talk of A.~Savin in these proceedings). DGLAP evolution 
would suppress the forward jet cross-section, for jets with $P_t^2\sim Q^2$ 
and 
low-$x$, because LO DGLAP evolution has strong $k_t$ ordering, 
from the target to the probe, and thus it 
cannot produce such events. The rate is also suppressed for NLO DGLAP. 
However BFKL evolution has no $k_t$ ordering 
and thus a larger cross-section for such events at both LO and NLO. 
The data do indeed show an 
enhancement of forward jet cross-sections wrt conventional NLO DGLAP 
calculations.
However this cannot be regarded as a definitive indication of the need for 
BFKL resummation because conventional calculations at higher order, 
$O(\alpha_s^3)$, do describe the data.

However, as we have already mentioned, even though conventional 
calculations do give reasonable fits to data, the peculiar behaviour of the 
low-$x$, low-$Q^2$ gluon gives us cause for some concern. Thorne and White 
have performed an NLL BFKL resummation and matched it to NLO DGLAP at high-$x$ 
in order to perform a global PDF fit. 
When this is done the gluon shape deduced from the 
scaling violations of $F_2$ is a lot more reasonable  
and a good fit is found to global DIS data, 
see the talk of C.White in these proceedings. A similar 
improvement to the gluon shape 
is got by introducing a non-linear term into the 
evolution equations, as done by Eskola et al~\cite{eskola}- but although this 
work has been widely used to give non-linear PDFs one must remember that it 
is limited to leading order.

\begin{figure}[tbp]
\vspace{-1.0cm} 
\centerline{
\epsfig{figure=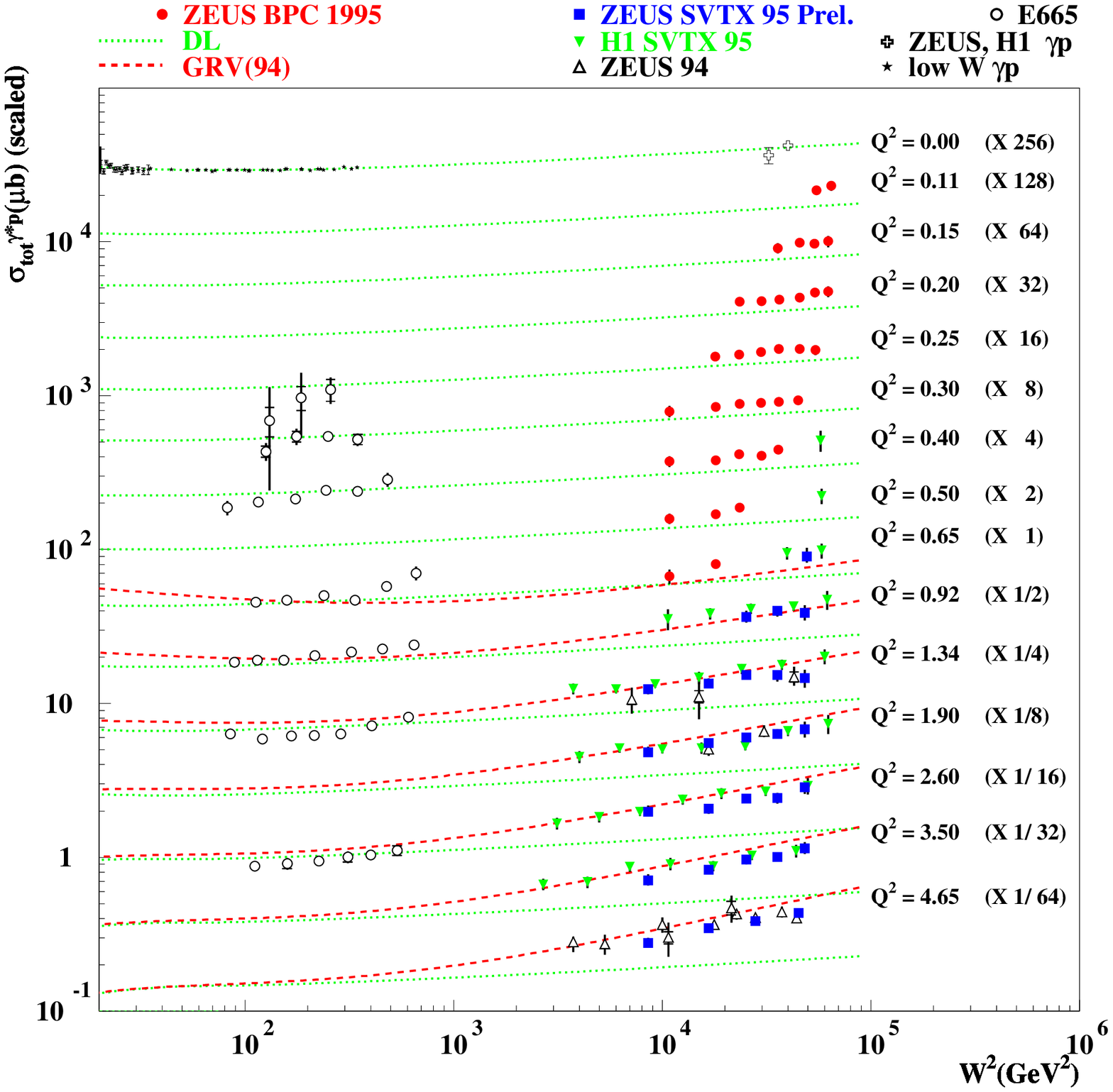,width=0.50\textwidth}
\epsfig{figure=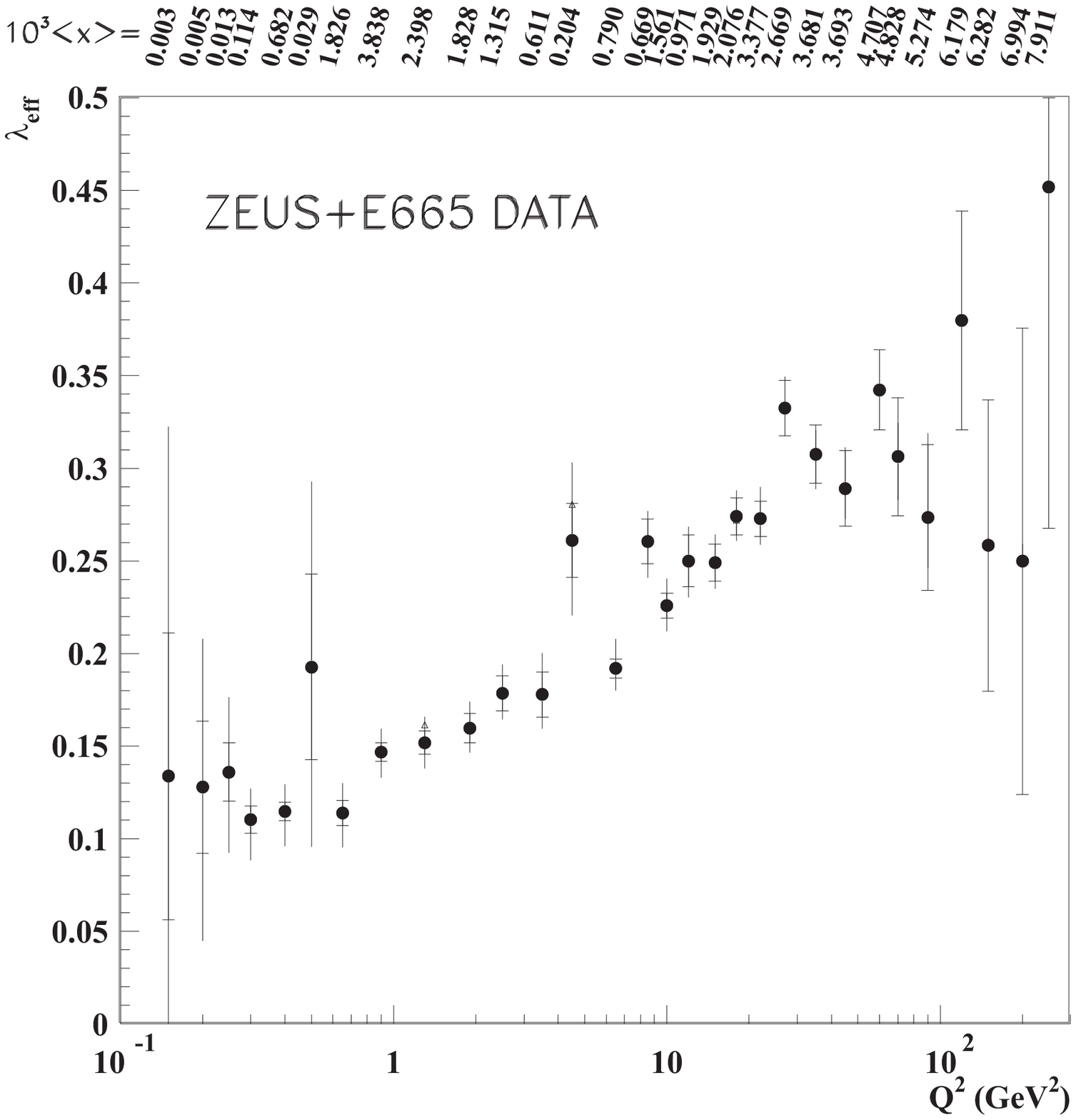,width=0.50\textwidth}
}
\caption {
Left plot: the photon-proton cross-section vs $W^2$ for various virtualities of the photon.
Right plot: the slope $\lambda= \partial lnF_2/\partial ln(1/x)$. 
}
\label{fig:siglam}
\end{figure}

These analyses make us suspect that the conventional formalism could be 
extended, but they are still not definitive. A different perspective comes 
from considering the low-$x$ structure function data in terms of the 
virtual-photon proton cross-section: at low-$x$, 
$\sigma(\gamma^* p)\sim 4\pi \alpha^2 F_2/Q^2$. 
The data are presented in this way in Fig.~\ref{fig:siglam} left-hand-side. 
A rise of $F_2(x) \sim x^{-\lambda}$, implies a rising cross-section 
with $W^2$, the centre-of mass energy of the photon-proton system, 
$\sigma(W^2) \sim (W^2)^\lambda$ (since $x=Q^2/W^2$ at low-$x$). However, the 
real-photon proton cross-section (and all high energy hadron-hadron 
cross-sections) rises slowly 
as $(W^2)^{\alpha-1}$, where, $\alpha=1.08$, is the 
intercept of the soft-Pomeron Regge trajectory. Thus the data on 
virtual-photon proton 
scattering are showing something new - a faster rise of cross-section than 
predicted by the soft-Pomeron which has served us well for many years. In 
Fig.~\ref{fig:siglam} right-hand-side we show the slope of this rise, 
$\lambda= (\alpha-1)$, as calculated from 
the data,  $\lambda = \partial lnF_2/\partial ln(1/x)$. One can see a change 
in behaviour at $Q^2 \sim 0.8$GeV$^2$ as we move out of the non-perturbative 
region 
-where the soft pomeron intercept gives a reasonable description of the data
 -to larger $Q^2$. Does this imply that we need a hard Pomeron as well? 

\begin{figure}[tbp]
\vspace{-2.0cm} 
\centerline{
\epsfig{figure=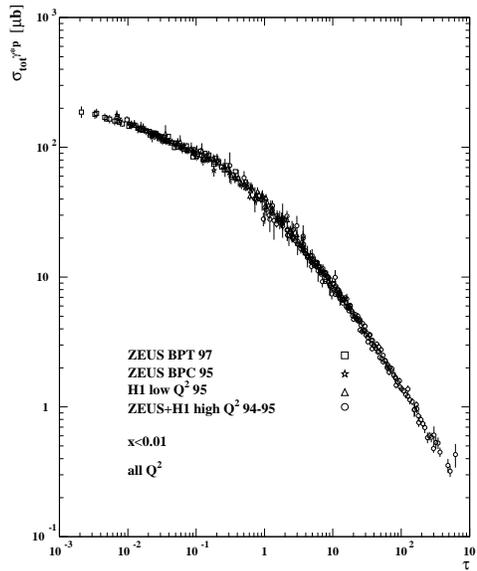,height=8cm}
}
\caption {$\sigma(\gamma^*p)$ vs the scaling variable $\tau=Q^2/Q_s^2$
}
\label{fig:geomscal}
\end{figure}

Dipole models have given us a way to look at  
virtual-photon proton scattering which can model the transition from the 
non-perturbative to the perturbative region. The interaction can be viewed as 
the virtual photon breaking up into a quark-antiquark pair and this pair, 
or dipole, then interacts with the proton. At low-$x$, the lifetime of the
$q\bar{q}$ pair is longer than the dipole-proton scattering time, such that the
physics is contained in the modelling of the dipole-hadron 
cross-section. There are many dipole models but the simplest Golec-Biernat 
Wusthoff model~\cite{golec} contains the essential features: 
$\sigma =\sigma_0 (1-exp(-r^2/(2R_0^2))$,
where $r$ is the transverse size of 
the dipole and $R_0$ is the 
transverse separation of the gluons in the target,
$R_0^2 =1/Q_0^2 (x/x_0)^\lambda$, where $x^\lambda \sim 1/(xg(x))$, 
is inverse to gluon density. Thus for small dipoles, 
$r < 1/Q$ and large $Q^2$,
 one obtains $\sigma \sim r^2 \propto 1/Q^2$ and Bjorken scaling 
(sophistications to the model correct this to give logarithmic 
scaling violation), 
whereas for large dipoles and small $Q^2$, one obtains $\sigma \sim \sigma_0$, 
ie a constant cross-section which corresponds to the correct photo-production 
limit. The reason that such dipole models have attracted attention in recent 
years is that the dipole-proton cross-section can be written in terms of a 
single scaling variable, $\tau$, $\sigma=\sigma_0 (1-exp(-1/\tau)$, where
$\tau=Q^2R_0^2 = Q^2/Q_0^2 (x/x_0)^\lambda$, rather than in terms of the two 
variables $x,Q^2$. This is known as geometric scaling, and evidence for it is 
shown by the low-$x$ ($x < 0.01$) data in Fig.~\ref{fig:geomscal}. 
Note that only low-$x$ data show this scaling. 
Geometrical scaling is predicted by many 
theoretical approaches to the low-$x$ regime which involve saturation and, 
$Q_s^2 = 1/R_0^2,$ is interpreted as a saturation scale below which 
non-linear dynamics applies. 

Note that the power $\lambda\sim0.3$, which describes the gluon density, 
$xg(x)\sim x^{-\lambda}$, within many dipole-models, is fitted to the data. 
It cannot be trivially related to the measured slope, 
$\partial lnF_2/\partial ln(1/x)$, at any $Q^2$, 
and it is not justified by the 
steep slopes of the gluon distribution observed at HERA- because such steep 
slopes are not in fact observed but are derived within the  
DGLAP formalism- which is explicitly not the formalism of most dipole models- 
and a steep slope $\lambda \geqsim 0.3$ is only found for 
$Q^2 \geqsim 10$GeV$^2$. However the saturation scale for 
HERA data is much lower, $Q_s^2\sim 1-2GeV^2$. 
Thus the steep slope of the gluon in the 
dipole models must be regarded as an input assumption. 

Geometric scaling is not unique to non-linear approaches, it 
can be derived from solutions to the linear 
BFKL equation~\cite{iancu} and even from the DGLAP equation~\cite{caola}. 
But note that such solutions 
do not extend into the low-$Q^2$ region and cannot give a picture of the 
transition from low to high-$Q^2$, as the dipole models do. Moreover, 
dipole models provide explanations for the constant ratio of the 
diffractive to the total cross-section data at HERA, and geometric 
scaling has also been observed in diffractive processes including 
vector meson production and 
deeply virtual compton scattering, see the talk of R.~Yoshida in these 
proceedings. These observations give hints that there is 
some truth to the dipole picture of saturation even though 
data at HERA are not definitive.

Even if the evidence for saturation at HERA is taken seriously 
the saturation scale is only, $Q_s^2 \sim 1-2$GeV$^2$, such that 
the region of non-linear dynamics largely coincides with the strongly-coupled 
region (where $\alpha_s$ is large). That is why there is interest in results 
from RHIC, where the nuclear environment enhances the high-density of the 
partons by $A^{1/3}$, such that saturation scales are higher, see the talk of 
A.~Dainese in these proceedings.
But what of the LHC?  Clearly ALICE data will be interesting, 
but even proton-proton data can be searched for signs of saturation if the
large rapidity region is considered, since small $x$ values are then 
accessed. For example, low-mass Drell-Yan data at LHCb can access 
$x \sim 10^{-6}$, see the talk of T. Shears in these proceedings.

If our conventional picture of DGLAP evolution in the HERA $x$ region is 
significantly
wrong then this will have implications even for classic Standard Model 
predictions, such as $W$ and $Z$ production in the central region of 
CMS and ATLAS. These bosons are produced at low-$x$, 
$5\times 10^{-4} < x < 5\times 10^{-2}$, in the central rapidity region, 
$-2.5 < y < 2.5$ and they are produced with enormous rate 
(even a modest 100 pb$^{-1}$ luminosity produces $10^6$ $W$ events) 
such that very early 
low luminosity running could show up discrepancies with our predictions.
Whereas rapidity spectra may not be much affected by unconventional $Q^2$ 
evolution~\cite{ball}, it should be fruitful to examine the boson $p_t$ 
spectra, since lack of $p_t$ ordering could affect these 
significantly~\cite{resbos}.

In summary, it is unclear that HERA data have actually given any evidence for
BFKL evolution, non-linear evolution or saturation, but there are hints in 
many places. The contribution of A.~deRoeck to this discussion considers the 
possibilities for further progress at HERA, the LHC and at future facilities.


\end{document}